# A NOVEL APPROACH FOR VERIFIABLE SECRET SHARING BY USING A ONE WAY HASH FUNCTION

**Keyur Parmar & Devesh Jinwala**

*Sardar Vallabhbhai National Institute of Technology, Surat*

*Email : keyur.beit@gmail.com & dcjinwala@gmail.com*

*Abstract*

*Threshold secret sharing schemes do not prevent any malicious behavior of the dealer or shareholders and so we need verifiable secret sharing, to detect and identify the cheaters, to achieve fair reconstruction of a secret. The problem of verifiable secret sharing is to verify the shares distributed by the dealer. A novel approach for verifiable secret sharing is presented in this paper where both the dealer and shareholders are not assumed to be honest. In this paper, we extend the term verifiable secret sharing to verify the shares, distributed by a dealer as well as shares submitted by shareholders for secret reconstruction, and to verify the reconstructed secret. Our proposed scheme uses a one way hash function and probabilistic homomorphic encryption function to provide verifiability and fair reconstruction of a secret.*

## 1 INTRODUCTION

Secret sharing, a famous term in cryptography, is a bit of misnomer. In a secret sharing, shares of the secret are distributed among a set of participants, not the entire secret. In cryptography, security of any algorithm is heavily dependent on the fact that an adversary not knowing the cryptographic key and so the key needs to be stored super-securely. The most secure way to safeguard the cryptographic key is to keep the key in a single well-guarded location but the idea is highly unreliable since a single misfortune can make the information permanently inaccessible. The obvious solution to safeguard the key from loss is to create backup copies but it increases the risk of security exposure. To enhance the security and reliability of the cryptographic keys, threshold secret sharing schemes were proposed independently by Adi Shamir [11] and G. Blakley [1] during 1979.

In a threshold secret sharing scheme, a dealer divides the secret, commonly a cryptographic key, into multiple parts called shares and distribute shares among a set of mutually suspicious participants in such a way that quorum of participants is required to reconstruct the original secret back; but fewer than the threshold number of participants can't recover any information about the secret. An interesting "real-world" example to explain this scenario was illustrated in Time Magazine. According to it, Russia used the "two-out-of-three" access control mechanism to control their nuclear weapons in the early 1990. Three parties, the President, the Defense Minister and the Defense Ministry, were involved to execute this scheme.

Shamir's threshold secret sharing scheme [11] has been extensively studied in the literatures to solve the problem of secret sharing. The Shamir's threshold secret sharing scheme is information theoretic secure but it does not provide any security against cheating as it assumes that the dealer and shareholders are honest but it is not always possible. A misbehaving dealer can distribute inconsistent shares to the participants or misbehaving shareholders can submit fake shares during reconstruction. To prevent such malicious behavior of cheaters, we need a verifiable secret sharing scheme, first introduced in 1985 by Benny Chor et al [3], through which the validity of shares distributed by the dealer are verified by shareholders without having any information about the secret. Initially, interactive verifiable secret sharing schemes were introduced but it requires interaction between the dealer and shareholders to verify the validity of shares and it imposes the enormous amount of extra overhead on the dealer, as it has to deal with a large number of shareholders. Later non-interactive verifiable secret sharing schemes were introduced to remove the extra overhead on the dealer, in which a share proves its own validity.

Existing approaches for verifiable secret sharing either verify the shares, distributed by a dealer or submitted by shareholders for secret reconstruction, or verify the reconstructed secret but not both. In order to verify shares, a dealer either transfers some additional information like check vectors or certificate vectors or it uses different encryption mechanisms. If existing verifiable secret sharing schemes are not using the check vectors or certificate vectors, security of such schemes depend on the intractability of computing hard to solve number theoretic problems in one way or another. If the scheme uses check vector or certificate vector then it increases an extra overhead on a dealer to compute and distribute that extra information among a large number of participants. In this paper, we use and extend the verifiable secret sharing approach to not only verify the validity of shares distributed by a dealer but to verify the shares submitted by shareholders for secret reconstruction, and to verify the reconstructed secret. Here, we use one way hash function and probabilistic homomorphic encryption function to prevent cheating done by a dealer or shareholders and to verify the reconstructed secret. We also assume a private communication channel between a dealer and each shareholder and broadcast channel among them in such a way that all participants receive the same broadcasted message.

The organization of the paper is as follows. The second section describes the related work to our topic verifiable secret sharing. In the third section, we explain preliminaries used throughout the paper. In the forth section, our proposed approach for verifiable secret sharing will be introduced and we will analyze it in the fifth section. Last section concludes the paper and references are at the end.

## 2  A SHORT SURVEY OF EXISTING SECRET SHARING SCHEMES

In a Shamir's secret sharing scheme [11], a dealer $D$ divides a secret $S$ into $n$ number of shares $S_1, S_2, ..., S_n$ and distributes the shares among a set of $n$ participants in such a way that for any threshold $t$, $t$ or more shares are able to reconstruct an original secret back but less then that leaves $S$ completely undetermined. Shamir's solution is information theoretic secure and does not rely on any unproven cryptographic assumption but it is not secure against cheating as it assumes that a dealer and shareholders are honest but it is not always possible.

The term verifiable secret sharing was first introduced by Benny Chor, et al. [3] to provide ability to an individual shareholder to verify his/her share distributed by a dealer. They proposed a constant round interactive verifiable secret sharing scheme and suggested a cryptographic solution, which is based on the intractability of factorization.

Then, Benaloh [4] proposed an interactive verifiable secret sharing scheme by using a homomorphism property and probabilistic encryption function. The scheme is only used to verify the shares distributed by the dealer but it does not verify, the shares submitted by shareholders for secret reconstruction and the reconstructed secret. Therefore, the scheme is only applicable in a situation where the shareholders are honest and the chance of accidentally submitting an erroneous share is zero. During the reconstruction, it assumes that the dealer is honest and reconstructs the secret correctly. In addition, the large number of encrypted shares of different polynomials increases the enormous amount of extra overhead on the dealer. As there are so much assumption about the honesty of a dealer and shareholders and the interactive nature of the scheme makes it infeasible to be used in a real world applications of secret sharing.

The first non-interactive verifiable secret sharing scheme was proposed by Paul Feldman [5] in which a share proves its own validity. The homomorphic relations exist between the values and their encryptions are utilized by the scheme but like other schemes, it has its own limitations. The scheme broadcasts an encrypted secret and so the security of the scheme depends on the hardness of inverting an encryption function. In addition, it does not verify the shares submitted by shareholders for secret reconstruction and it does not verify the reconstructed secret.

Recently published paper in verifiable secret sharing, proposed by Lein Harn and Changlu Lin [7], utilizes the redundant shares to detect and identify the cheaters. In their scheme, they assume that a dealer is honest. In their approach, shareholders only try to verify the reconstructed secret by using the majority of secret mechanism. The scheme only detects and identifies the cheaters when more than the threshold number of shares are available and majority of them and more than the threshold number of shares are from the honest shareholders. When there are only threshold number of shares are present, the scheme works exactly same as Shamir's [11] scheme and does not detect the cheating.

In 2008, Zhengjun Cao, et al. [2] proposed a scheme by using a hash function to verify the validity of the reconstructed secret. They have revisited the approach of Obana, et al. [8], proposed at Asiacrypt'2006 and gave a more efficient solution to the problem of verifiable secret sharing. In their scheme, they can only verify the reconstructed secret. They assume that the dealer is honest and does not distribute any fake shares. Also, the scheme assumes that the shareholders receive the correct shares of the secret without any exception. At the end, the scheme is able to verify the reconstructed secret but if it is not correct then it cannot identify the faked shares and cheaters. In a real world, after the reconstruction of a secret, the secret is used to trigger some event or to perform some action and if it is not correct then it fails to do so. Therefore, we can verify the correctness of a secret at that time also, as we reconstruct the secret only when we need it. The scheme fails when the dealer or shareholders are not honest and intentionally or accidentally use fake shares during the secret sharing.

## 3  PRELIMINARIES

### 3.1  Definition of Threshold Secret Sharing Scheme

Divide some secret $S$ into $n$ pieces $S_1, S_2, ..., S_n$ and distribute them among a set of $n$ shareholders in such a way that for any threshold value $t$, the knowledge of any $t$ or more $S_i$ pieces makes $S$ easily computable but the knowledge of any $t-1$ or fewer $S_i$ pieces leaves $S$ completely undetermined. Such a scheme is called a $(t, n)$ - threshold secret sharing scheme [11]. Here, $1 \leq i \leq n$.

### 3.2  Probabilistic Encryption Functions

The encryption function used in our proposed approach is based on an idea of probabilistic encryption [5] [4] [6] found in [6]. A deterministic encryption scheme always produces the same ciphertext for a given pair of plaintext and key and so cryptanalyst gets some information every time when he/she encrypts the plaintext. Probabilistic encryption was invented to eliminate the information leakage with a public key cryptosystem. The goal is to stop the cryptanalyst to extract any information about the plaintext from a corresponding ciphertext. In probabilistic encryption, an encryption algorithm each time generates a different ciphertext for a same pair of plaintext and key. If cryptanalyst has a plaintext, encryption key, encryption algorithm and ciphertext, he cannot be sure that the given ciphertext was the encryption of his plaintext or not.

To develop an encryption function $E$, choose large and distinct primes $p$ and $q$. Let $N$ be the product of $p$ and $q$. Select a large prime $r$, greater than the size of the secret domain. Find some quadratic non-residue $y$ such that, the Legendre symbols satisfy

$$\left(\frac{y}{p}\right) = \left(\frac{y}{q}\right) = -1$$

Hence, the Jacobi symbol $\left(\frac{y}{N}\right)$ is +1. The public key pair is $(N, y)$; the private key pair is $(p, q)$. To use $E$ to encrypt a share $S$, the dealer $D$ randomly selects $x \in \mathbb{Z}_n^*$ and outputs the value $x^r y^s \bmod N$. Trapdoor factors $(p, q)$ of $N$ must be needed to recover the original value $S$ from its encrypted value.

### 3.3 Homomorphism Property and Homomorphic Encryption

Let, in any $(t, n)$ - secret sharing scheme, $F_I$ is a function that reconstructs the secret $S$ from any subset of $t$ or more shares where $I \subseteq \{1, 2, \ldots, n\}$ and $|I| \geq t$, where $t$ is a pre specified threshold value.

$$S = F_I(S_1, S_2, S_3, \ldots, S_t)$$

Let $\oplus$ and $\otimes$ are any two binary operators defined on the elements of a secret domain and a share domain respectively. If $(t, n)$ secret sharing scheme has the $(\oplus, \otimes)$- homomorphism property [5] [4] then for all $I$, whenever

$$S = F_I(S_1, S_2, S_3, \ldots, S_t) \text{ and } S' = F_I(S_1', S_2', S_3', \ldots, S_t'),$$

Then we easily compute,

$$S \oplus S' = F_I\left(S_1 \otimes S_1', S_2 \otimes S_2', S_3 \otimes S_3', \ldots, S_t \otimes S_t'\right)$$

Homomorphism property implies that the compositions of shares of the secrets are shares of the composition of the secrets. Shamir's polynomial based $(t, n)$ – secret sharing scheme is $(+, +)$ - homomorphic and so the sum of shares of the secrets are shares of the sum of the secrets.

In homomorphic encryption, the homomorphism property provides the ability to perform computations on the ciphertext without decrypting it. The encryption function $E$ is homomorphic if given $E(x)$ and $E(y)$, one can obtain $E(x \oplus y)$ without decrypting $x$, $y$, for some operation $\oplus$. In our $(t,n)$ - secret sharing scheme, we broadcast the encrypted values of coefficients and so each shareholder can construct an encrypted polynomial and compute the encrypted share value for different shareholders. Hence, shareholder can compare his encrypted share value with the encrypted value retrieved from the polynomial. This property is used in our scheme to verify that the shares are collectively $t$-consistent in such a way that every subset of the threshold number of shares can reconstruct the same secret.

### 3.4 One Way Hash Function

A one-way hash function[8][2] takes a variable length input message $M$ and converts it into a fixed-length output $H(M)$ called hash code or message digest, such that the following properties hold.

1. For any hash code $h$, it is computationally infeasible to find $x$ such that $H(x) = h$.

2. For any given message $M_1$, it is difficult to find another message $M_2 \neq M_1$, such that $H(M_1) = H(M_2)$.

3. It is difficult to find two random messages, $M_1$ and $M_2$, such that $H(M_1) = H(M_2)$.

## 4 OUR PROPOSED APPROACH FOR VERIFIABLE SECRET SHARING

### 4.1 Share Generation and Distribution Phase

Input: Secret $S \in GF(p)$ and a public hash function $H$.

Output: Shares of the secret $S$, $S_i$ where $i = 1, 2, 3, \ldots, n$.

1 Dealer $D$ chooses a large prime $p > \max(S, n)$.

2 Then it selects $t$-1 random independent coefficients, $a_1, a_2, a_3, \ldots, a_{t-1}$ where, $0 \leq a_i \leq p - 1$.

3 Select the random polynomial and set, $a_0 = S$.

$$f(x) = a_0 + a_1 x + a_2 x^2 + \ldots + a_{t-1} x^{t-1}$$

4 Compute the share of the secret for each shareholder and distribute the pair $(i, S_i)$ to each shareholders.

$$S_i = f(i) \text{ Where, } 1 \leq i \leq n$$

5 Dealer uses one way hash function to generate hash code for every share,

$$H(S_i) \quad \text{Where, } 1 \leq i \leq n$$

6. Dealer broadcasts, encryption of all the $t$ coefficients $E(a_0), E(a_1), ..., E(a_{t-1})$ and hash code of the secret $H(S)$.

7. Dealer maintains a public file for hash code of every share as well as hash code of the secret

8. Each $i^{th}$ shareholder verifies his/her share as follows: Shareholder uses same one way hash function to generate hash code and compare it with the hash code available in the public file and also computes an equation,

$$E(f(i)) = E(a_o) \oplus (E(a_1) \otimes E(i^1)) \oplus (E(a_2) \otimes E(i^2))...\oplus (E(a_{t-1}) \otimes E(i^{t-1}))$$

9. If this equation holds, the shareholder accepts his/her share as correct.

10. If all the shareholders find their shares correct, then only the dealing phase is completed successfully. Dealer discards $S, a_1, a_2, ..., a_{t-1}$ and decryption key, the information needed to decrypt the encrypted values

11. Otherwise, it is up to the honest shareholders to decide whether it is the Dealer or the accuser that misbehaves.

### 4.2 Share Reconstruction Phase

Input: Shares $S_i$ Where $i \subseteq \{1, 2, 3, ..., n\}$ and $|i| \geq t$, a public hash function $H$

Output: Secret $S$

1. Dealer verifies each share by generating hash code for each share and then compares it with the corresponding hash code available in the public file and accept it only if they are equal.

2. If $t$ or more than $t$ shares are available then the dealer computes an interpolated polynomial $f(x)$ at $t$ or more points $(i_1, S_1), (i_2, S_2), ..., (i_t, S_t)$.

3. Here, the constant term of a polynomial is the secret $S$.

4. A dealer as well as shareholders, verify the secret by generating hash code for secret $S$ and then comparing it with the Hash Code available in a public file.

## 5  ANALYSIS OF OUR ALGORITHM

In our algorithm, we extend the Shamir's original threshold secret sharing scheme [11] to verify the shares and the secret. We choose the random polynomial in $GF(p)$ where the coefficients are also chosen randomly in $GF(p)$. We set the secret as a constant term of the polynomial. Now, we can use the polynomial to generate the shares of a secret and distribute it among a set of shareholders. Up to this point, our scheme works same as Shamir's scheme [11].

Here, one way hash function is used to produce a hash code for a secret as well as for shares, distributed among shareholders. A dealer maintains a public file for a hash code of each share as well as for a hash code of the secret. Here, we assume that the one way hash function is available to all the shareholders but its one way property ensures that it is infeasible to reconstruct the share from its hash code.

In our algorithm, each shareholder verifies his/her share by computing a hash code of the share and then comparing the hash code with the hash code available in the dealer's public file. If they are not equal then shareholder gets an assurance that the erroneous share is received or the malicious intruder has modified the original share distributed by the dealer and shareholder can contact a dealer for his correct share. If both the hash codes for a share, computed by a shareholder and available in a public file, are equal then each shareholder can check the consistency of his/her share using the probabilistic homomorphic encryption function. Dealer has transmitted the encrypted coefficients of a polynomial and the encryption function is probabilistic, and so each shareholder encrypts his/her share and compute the encrypted polynomial to verify his/her share. By using the homomorphic nature of an encryption function, the polynomial computed with an original values and then encrypted or the polynomial computed with encrypted coefficients and share, both are equal. Here the encrypted constant coefficient of the polynomial is our original secret, and so we can say that our share is generated from original secret and it is consistence with other shares generated from the same secret by a same polynomial. Now, each shareholder can obtain high confidence that he/she holds a valid share of the secret rather than a useless random number.

At the end of a share generation and distribution phase, a dealer discards the original secret, coefficients of the polynomial, and the share values that is distributed to different shareholders and the information needed to decrypt the encrypted values. Here, the decryption key is permanently removed so that nobody can ever decrypt those encrypted coefficients where the constant coefficient of the polynomial is our original secret. After the end of this phase, each shareholder possesses a valid and consistent share of a secret.

During the reconstruction phase, each shareholder submits a share to the dealer and dealer verifies the share by computing the hash code for it and then comparing it with the hash code available in the public file.

If both values are equal then accept it as a valid share otherwise reject it. If there are more than a threshold number of shares available for secret reconstruction, then the dealer can reconstruct the original secret back by using any polynomial interpolation technique like Lagrange's interpolation. Reconstructed secret is verified by computing the hash code for it and comparing it with the hash code available in a public file. If both of them are equal then the reconstructed secret is correct.

## 6 CONCLUSION

Despite the previously mentioned flaws with verifiable secret sharing, at least a heavily modified version was needed to use the secret sharing in a cryptographic key management. A novel approach for verifiable secret sharing is presented in this paper for a fair reconstruction of the secret. In our verifiable threshold secret sharing scheme, each shareholder verifies that the share comes from an honest dealer and it is not an erroneous piece or it is not modified by a malicious intruder. The shareholder also verifies consistency of a share by using the probabilistic homomorphic encryption function in such a way that each shareholder gets an assurance that the share is a valid piece of the secret and when clubbed with other $t$-1 shares, it will reconstruct the original secret back. During the reconstruction, the dealer verifies each share by using the one way hash function and accepts it if it is a valid share. If the dealer has more than a threshold number of shares available then he/she can reconstruct the original secret back. The reconstructed secret is verified by a dealer and shareholders by using a one way hash function. Our proposed scheme is used in a situation where the dealer and/or shareholders are not honest. We use it to detect and prevent the cheating as well as identify the cheaters and it can be used in a situation where the mutually suspicious participants should collaborate in a secure and reliable way.